# Anisotropic Structure of the Order Parameter in FeSe$_{0.45}$Te$_{0.55}$ Revealed by Angle Resolved Specific Heat

B. Zeng*, G. Mu*, H. Q. Luo*, T. Xiang*, H. Yang*, L. Shan*, C. Ren*, I. I. Mazin†, P. C. Dai*‡ & H.-H. Wen*

*National Laboratory for Superconductivity, Institute of Physics and National Laboratory for Condensed Matter Physics, Chinese Academy of Sciences, Beijing 100190, China

†Code 6391, Naval Research Laboratory, Washington, DC 20375, USA

‡ Department of Physics and Astronomy, The University of Tennessee, Knoxville, Tennessee 37996-1200, USA

**The symmetry and structure of the superconducting gap in the Fe-based superconductors are the central issue for understanding these novel materials. So far the experimental data and theoretical models have been highly controversial. Some experiments favor two or more constant or nearly-constant gaps, others indicate strong anisotropy and yet others suggest gap zeros ("nodes"). Theoretical models also vary, suggesting that the absence or presence of the nodes depends quantitatively on the model parameters. An opinion that has gained substantial currency is that the gap structure, unlike all other known superconductors, including cuprates, may be different in different compounds within the same family. A unique method for addressing this**

issue, one of the very few methods that are bulk and angle-resolved, calls for measuring the electronic specific heat in a rotating magnetic field, as a function of field orientation with respect to the crystallographic axes. In this Communication we present the first such measurement for an Fe-based high-$T_c$ superconductor (FeBSC). We observed a fourfold oscillation of the specific heat as a function of the in-plane magnetic field direction, which allowed us to identify the locations of the gap minima (or nodes) on the Fermi surface. Our results are consistent with the expectations of an extended s-wave model with a significant gap anisotropy on the electron pockets and the gap minima along the ΓM (or Fe-Fe bond) direction.

The discovery of high-temperature superconductivity in Fe pnictides[1] was arguably the most important event in the physics of superconductivity since the discovery of cuprate superconductors. Despite an impressive progress in the last two years, the most basic question: what is the symmetry and structure of the order parameter? – remains unanswered. Yet, without firmly answering this question further theoretical advance, most importantly, identification of the pairing mechanism, becomes essentially impossible.

It would be unfair to say that we are entirely in the dark (see reviews [2,3] for detailed discussions). Indeed, it seems highly unlikely that very similar materials with the same main element, the FeAs or FeSe planes, would have qualitatively different pairing symmetries (quantitative differences are of course possible). For several members of the family of Fe-based superconductors the NMR data have positively identified the parity of the superconducting state as singlet, essentially leaving only two possibilities for the angular momentum of the Cooper pairs: L = 0 (s-wave) and L = 2 (d-wave). These can come in several varieties. Specifically, in the tetragonal symmetry several symmetry-distinguishable versions of the d-wave pairing are allowed, with the basis functions transforming as $x^2$-$y^2$, $xy$, and $xz$, $yz$ or $xz±yz$ (we exclude the chiral $xz±iyz$ on the basis of the μSR experiments that would have detected a spontaneous magnetization below $T_c$ in a chiral state).

Furthermore, substantial indirect evidence has been accumulated against the relevant d-wave symmetries: (1) the Josephson current is finite and sizeable along the $z$ direction in some of the FeBSCs.[4] This excludes any order parameter that integrates to zero over $x$ and $y$ at every given $z$, including the relevant *d*-wave symmetries; (2) Angular resolved photoemission (ARPES) shows that in many FeBSC (1111, 122 and 111 series) the gap is fairly isotropic on each Fermi surface pocket.[5,6] This, again, excludes $x^2$-$y^2$ and $xy$ near the surface of the samples, and, by implication, in the bulk. Preliminary data exist also for a 11 material similar in composition to ours, although these data are still incomplete as regards the electron-like Fermi surface.[7] (3) Absence of the Wohlleben (paramagnetic Meissner effect) in some 1111 materials indicates that weak links in polycrystalline samples have predominantly 0 phase shifts, while both $x^2$-$y^2$ and $xy$ symmetries imply roughly 50:50 distributions of 0 and $\pi$ phase shifts.[8]

At the same time, multiple evidence indicates existence of unpaired quasiparticles at small energies, sometimes all the way down to nearly zero energies.[9,10] The most natural explanation would be that the excitation gap has nodes somewhere on the Fermi surface, and the nodal areas provide low-energy quasiparticles. Yet another explanation, quite popular now, is that the low-energy excitation are due to the pair-breaking effects of defects and impurities in the scenario of $s_\pm$ pairing.[11] Arguably the most sensitive probe of the low-energy excitation structure, thermal conductivity and penetration depth measurements, seem to point to nodal or nearly-nodal states in three FeBSC, $KFe_2As_2$[12], $LaFePO$[13,14], and $BaFe_2(As,P)_2$,[15] and to full gaps with various degrees of anisotropy in many others.[16-19] It was suggested, based on the anisotropy of the penetration depth, that some materials may have a horizontal nodal line consistent with the $xz\pm iyz$ d-wave state (which however is unlikely for other reasons), or with an accidental nearly-horizontal nodal line within an extended s-wave model.[20]

At this point, it becomes truly indispensable to address the angular structure of the order parameter directly, and in the bulk. Indeed, as explained below, in this way one can distinguish between an isotropic band or a horizontal nodal line, on one hand, and a vertical nodal line, on the other. If other experiments (*e.g.*, thermal conductivity) suggest presence of nodes, *and* there is no *ab* plane angular dependence, this indicates a horizontal nodal line. If, however, an angular dependence is detected, this demonstrates the presence of vertical nodes, and pinpoints their location on the Fermi surface. If, finally, such a dependence is present, but angle-integrated experiments exclude nodes, one concludes that the material has a highly anisotropic nodeless order parameter (with known location of gap minima). This is invaluable information, but how can it be harvested?

The most effective and accurate techniques for addressing the angular structure of the order parameter rely on probing thermodynamical properties in a rotating in-plane magnetic field. Recently, Matsuda et al. [21] have measured the thermal conductivity in such a field in $BaFe_2(As_{1-x}P_x)_2$, where the zero-temperature, zero-field limit of the same quantity suggests the presence of gap nodes. They concluded that such nodes exist on the electronic Fermi surface, centered around the M point.

In this paper we have used a different approach, Angle Resolved Specific Heat (ARSH) measurements, and we have selected a material for which thermal conductivity at low temperature indicates an anisotropic order parameter without nodes[22], a representative of the so-called 11 family, $FeSe_xTe_{1-x}$. Indeed we find a strong fourfold anisotropy in ARSH, however different from that suggested in Ref.21 for $BaFe_2(As_{1-x}P_x)_2$ from their thermal conductivity data.[21] Given the result of thermal conductivity of Dong et al.,[22] we speculate that FeSe$_x$Te$_{1-x}$ has a strongly anisotropic s-wave nodeless gap, with deep minima along the ΓM (the Fe-Fe bond in the real space) directions. It seems difficult to reconcile the evidence of Ref. 22, and our results, both indicating deep minima, with the tunneling data of Hanaguri et al. [23] who have observed a full gap of about 1.4 meV with no detectable subgap DOS in the same compound. Below we will argue that such a reconciliation is nevertheless possible.

On the theory side, most attempts to deduce the angular structure of the order parameter from model calculations on 1111 and 122 materials roughly agree that deep minima or gap nodes appear, in some relevant parameter range (see. Ref. [24] and references therein) usually along the ΓM direction on the outer electron Fermi surface. Chubukov and Eremin [24] have pointed out that if the order parameter can be expanded in spherical harmonics around the M point, then in the lowest order, the gap minima can only occur along the ΓM direction, either on the outer or on the inner surface.

The FeSe$_{0.45}$Te$_{0.55}$ single crystals were grown using the self-flux method. Details regarding the growth of the samples are given in the Supplementary Materials (Method-I). The diamagnetic signal in zero-field-cooling (ZFC) demonstrates a perfect Meissner effect, as shown in the main panel of Figure 1. In the inset (a) of the Figure-1 we plot the resistivity in a wide temperature range, a broad maximum of resistivity appears in the middle temperature regime, which seems to be an intrinsic feature of the FeSe$_x$Te$_{1-x}$ system and has been reported by other groups.[25] Note a very sharp resistive transition at about $T_{c,mid}$ = 14.5 K with a transition width narrower than 0.5 K (10-90% $\rho_n$). In the inset (b) we show the temperature dependence of the upper critical fields with two field orientations: H∥ab and H∥c. It is clear that the superconductivity is very robust against the magnetic field. In H∥c, the superconducting transition temperature drops by about 1.79 K at 9 T, but only by 0.91 K in H∥ab. The slope of the upper critical field is $dH_{c2,ab}(T)/dT|_{T=Tc}$ = -10.4 T/K, and $dH_{c2,c}(T)/dT|_{T=Tc}$ = -5.26 T/K, yielding an anisotropy of about 2. The resistance measurements are described in detail in the Supplementary Materials (Method-I). The sharp magnetic and resistive transitions, together with the very small residual specific heat coefficient $\gamma_0$ (see below) demonstrate the good quality of the sample and allow us to proceed with the angle resolved specific heat (ARSH) measurements.

In a type-II superconductor, applying a magnetic field (shown by the thick arrow in the Fig4sup of Supplementary (Method-III)) induces vortices extending along the field direction, with a supercurrent flowing in the perpendicular plane, as shown by the red arrow and circle in the Fig4sup. Due to the motion of this flowing electron condensate, the local quasiparticle density of states (QP DOS), $N(\varepsilon)$, in the area surrounding the vortex core will be affected by a Doppler shift with the energy $\delta E = m\mathbf{v}_s \cdot \mathbf{v}_F$, as $N(\varepsilon) = N(\varepsilon_0 \pm \delta E)$, with $\delta E$ for the band number (i) given by [26-28]

$$\delta E^{(i)} = m\mathbf{v_s} \cdot \mathbf{v}_F^{(i)} = \frac{E_H^{(i)}}{\rho}(\widetilde{v}_{F,y}^{(i)} \cos\alpha - \widetilde{v}_{F,x}^{(i)} \sin\alpha)\sin\omega. \qquad (1)$$

Here α is the angle between the external field and the Fe-Se-Fe bond (45° away from the ΓM direction), ω is the winding angle of the supercurrent around the vortex, and $\widetilde{v}_{F,x(y)}^{(i)}$ is the normalized (to the average value of the Fermi velocity in the corresponding band), $\langle v_F^{(i)} \rangle$, x-(y-)component of the Fermi velocity. The characteristic Doppler shift energy scale $E_H^{(i)}$ is defined as [28] $E_H^{(i)} = a\hbar \langle v_F^{(i)} \rangle \sqrt{2H/\pi\Phi_0}$, where $\Phi_0$ is the flux quantum, and $a$ is a geometrical factor taking 0.465 for the triangle vortex lattice; $\rho$ is a dimensionless variable characterizing the distance from the vortex core. This energy shift, as discussed for instance by Graser et al [29], leads to the energy shift of the DOS curve, and thus, for a clean BCS superconductor, to the following equation for the DOS at the Fermi level (in units of the normal DOS):

$$N^{(i)}(0) = \text{Re} \left\langle \left\langle \frac{|\delta E^{(i)}|}{\sqrt{|\delta E^{(i)}|^2 - \Delta_k^{(i)2}\rho^2}} \right\rangle_H \right\rangle_{FS}, \qquad (2)$$

where the first averaging is performed over the unit cell of the vortex lattice and the second over the Fermi surface. For a nodal or quasi-nodal superconductor the main **k** dependence comes from the order parameter in the denominator, therefore we have kept this subscript explicitly in $\Delta$. The maximal value of $\rho$ is 1 [29], therefore, in principle, in a nodal superconductor there will always be directions near the nodes where N(0) is nonzero (where $|\delta E^{(i)}| \geq \Delta^{(i)}_k$), reaching its normal value right at the nodal points. An important thing to appreciate is that there is always some residual DOS at the nodes, which is simply enhanced by the effect of the magnetic field, and the net effect of this enhancement is the strongest when all nodes are "excited", and not just a few of them. This leads to a slightly counterintuitive result that N(0) is maximal when the magnetic field is aligned with an antinodal (and not a nodal line).

It should be pointed out that this result holds only for sufficiently low temperatures and fields. For a given magnetic field, there is a reversal temperature, $T^*$, and at higher temperature N(0) is maximal when the magnetic field is aligned with an nodal (and not an anti-nodal line). This result appears to be quite universal: it was obtained in the Doppler-shift approximation [30], as well as in more accurate approaches [31,32,33]. The exact value of $T^*$, however, is rather model-sensitive and the only way to determine in which regime one actually works is to repeat the measurements at different temperatures and until the actual reversal has been obtained. This has been done, for instance, in Ref. [34] for CeCoIn$_5$, and we also have found a reversal as a function of temperature in our experiment. This allowed us to pinpoint the positions of the gap minima with complete certainty.

As opposed to the position of the minima, the detailed shape of the oscillation, albeit formally dependent on the gap structure, is much harder to determine; it is important to remember that even in the most favorable cases disorder or 3D dispersion considerably reduce the amplitude of the oscillations compared to the original calculated value for a 2D d-wave gap (10% is usually considered to be a reasonable upper limit for realistic systems).[26,27] In fact, the ARSH has been successfully applied in identifying the nodal gap structure in such superconductors as CeRhIn$_5$ [35] and YNi$_2$B$_2$C[36] (in both cases about 5% oscillations have been detected). Note that in a "quasinodal" case, when the gap does not have true nodes, but has a very small value along particular directions, the discussion above still holds, as long as $E_H^{(i)} > \Delta_{\min}^{(i)}$.

In Figure 2(a)-(c) we present our experimental results for FeSe$_{0.45}$Te$_{0.55}$ single crystals. We rotate the sample with the FeSe plane parallel to the direction of a fixed magnetic field and measure the variation of the specific heat. As shown in Fig.2(a)-(b), the specific heat coefficient $\gamma = C/T$ shows clear fourfold oscillations in the in-plane magnetic field of 9 T. At the low temperatures T = 2.6-2.7 K, the minimum appears when the field is along the ΓM direction. While when the temperature is increased to about T = 3.65-3.75 K, the oscillations are still visible, but are reversed with the maxima along the ΓM directions (see Fig.2(a)). This is extremely important, because recent theoretical estimates have already predicted the reversal effect for iron-pnictides [24,37] and placed the reversal temperature, $T^*$, at a rather low temperature, while our data unambiguously prove that 2.7 K < $T^*$ < 3.6 K. In order to show that, in Fig.2(c), we present the difference of C/T between the data collected at α = 0° and 45°. One can see a clear crossover between the two regions at about 2.9 K at H = 9 T. Our data is quite consistent with the theoretical prediction of the reversal effect. Because of that, we can say with complete certainty, that the gap minima are located along the ΓM lines.

As an additional test, we have also measured the SH of the same sample in H = 0, as well as in an isotropic superconductor Nb in H = 0.4 T while rotating the sample. As shown in the Supplementary Materials (Method-II), within the error bars, we have not seen any oscillations of ARSH in either case. Therefore we believe that the oscillations observed in the in-plane field of 9 T for FeSe$_{0.45}$Te$_{0.55}$ (Figure 2) really reflect the gap structure in this material.

As discussed above, the observation of the oscillations of SH tells us that the minimum gap should be smaller than the Doppler shift energy $E_H$. Our LDA calculations yield the average Fermi velocity $\tilde{v}_F^{ab}$ that is similar for the two hole sheets and the outer electron sheet, varying between 2.5×10$^5$ and 2.6×10$^5$ m/s, and $\tilde{v}_F^{ab}$ ≈ 4.0×10$^5$ m/s for the inner electron sheet. Assuming a many-body renormalization of the order of 2--3, and H = 9 T, we estimate $E_H$ to be between 1 and 2 meV. Thus, these results present a direct bulk evidence of a strong gap anisotropy with nodes or deep minima. Now we consider the Fermi surface topology in this system. Calculations [38] and ARPES [7] indicate the same generic

Fermiology as in Fe pnictides, which includes two sets of the Fermi surfaces: the hole one contains two or three roughly cylindrical pockets around the Γ point at the center of the Brillouin zone (0,0), and the electron one consists of two overlapping warped elliptical cylinders near the M point (π,π) points. The points near which the gaps nodes can be located, according to our ARSH data, are shown by the colored markers in Figure 2(d). Note that the ellipticity in $FeSe_xTe_{1-x}$ is rather small compared, say, to 122 materials, so the effects of the FS and the Fermi velocity anisotropy should be very weak (the calculated Fermi surfaces are shown in Figure 3(b)-(c); in the cartoon representation of Figures 2(d) the ellipticity is intentionally exaggerated). Therefore we conclude that the oscillations of specific heat in our samples are not due to the anisotropic Fermi velocity.[31-32] This is corroborated by the fact that the fourfold oscillations of the specific heat reverse their sign when the temperature is changed slightly. The anisotropy of the Fermi velocity cannot change appreciably when the temperature is increased just from 2.6 K to 3.7 K.

A canonical $x^2$-$y^2$ (*xy* in the one-Fe unit cell) d-wave pairing on the hole-FS would readily explain our data (shown by the red lines in Figures 2(a) and (b)). Indeed, it would have created the nodal lines at the right spots in both hole bands, and electron pockets (red and blue balls, and red dashes in Figure 2d). However, as discussed in the introduction, there is convincing evidence against this pairing state. Another viable alternative is that the nodal lines form on the electron FSs, as indicated also from recent Raman scattering measurements,[39,40] at the points marked by the red and blue balls on the electron pockets (Figure 2(d)).

Indeed a number of model calculations [41-43] predict nodes at the electron FS in some parameter range. The gap structure in these calculations is set by the orbital composition of the electron bands in the unfolded Brillouin zone, and nodes appear roughly where the character changes from the *xy/(x²-y²)* orbitals to the *xz/yz* orbitals. In Figure 2(a)-(b) we compare simulated curves with the nodes located at the ΓM line, α=45° (*e.g.*, d-wave) with the experiment (see the Supplementary Material Method-III for the description of simulations). We observe that this scenario is consistent with the experiment, as would an extended *s-wave* model with the nodes located within ±5° from 45° (not shown here). However, nodes that are farther away from α = 45° do not describe the data well enough. On the other hand, existing calculations [43] suggest that for the parameters appropriate for FeSe no nodes are present at all. Thermal conductivity [22] indicates that there are no zero nodes in Fe(Se,Te) superconductors, but deep minima. All in all, this points to a scenario with no nodes, but deep minima on either inner or outer electron barrels, or both, as shown by the blue or red balls in Figure 2(d).

Assuming that the order parameter varies smoothly with the angle, one can expand it as a function of winding angle on the relevant Fermi surface[24,37]. If, as suggested by model calculations, the anisotropic part is in the gap on the electron pockets, we can expand it as[24], $\Delta_e(\varphi) = \Delta_0[1 - r\sin 2\varphi]$ where the winding angle φ is counted with respect to the Fe-Se-Fe direction, as shown in Fig.2(d), and *r* controls the gap anisotropy. In Figure 3(a) we show the angle dependence of the gap on the electron pocket centered at M(π,π), taking $\Delta_0$ = 1.3 meV. For |*r*|<1 there are no nodes, but minima at φ = ±45°, located on the outer (*r*>0) or inner (*r*<0) barrels. For |*r*| > 1, pairs of nodes appear on both sides of these directions and they are moving away as the anisotropy parameter *r* grows.

Several notes are in place here: (i) model calculations for 122 compounds suggest $r>1$ [42,43]; (ii) a model calculation for FeSe suggests $0 < r < 1$ (in fact, $r\sim0.6$)[43] (iii) thermal conductivity suggests no nodes, that is, $|r| < 1$ [22] and (iv) STM suggests an isotropic gap ($r \sim 0$)[23]. It is worth noting that from the point of view of our experiment we cannot distinguish the sign of $r$: nodes or minima on the inner barrel or on the outer barrel would produce practically the same ARSH spectra. Finally, since the oscillations can only be observed when the Doppler shift energy is larger than or comparable to the minimum value of the gap, the gap minima should be around 1 meV or less. Using the gap value estimated from STM, 1.4 meV, we find that $|r| > 0.25$.

In principle, our data are consistent with the Functional renormalization-group calculations of Ref. [43] (and it is reasonable to assume that random phase approximation (RPA) calculations would give similar results), as well as with thermal conductivity measurements, suggesting that $r$ is large, but not larger than 1. However, STM experiments bring in a new dimension since they do not detect any subgap density of states at all. Interestingly, it is possible to reconcile *all* experimental results among themselves (but not with the model calculations) if we assume that $r$ is large but *negative,* and take into account the fact that the STM current is dominated by those part of the Fermi surface that have sizeable Fermi velocity along the tunneling direction (crystallographic c-axis). As Fig.3(c) shows, the Fermi velocity along c-axis is not negligible only for the outer barrel, while for the inner barrel (and for both hole Fermi surfaces, for that matter), it is vanishingly small. Thus, if the minima are on the *inner* barrel, they would not be seen in the STM experiment, but would be in ours, and, interestingly, in the thermal conductivity experiment: as Fig. 3(b) shows, the inner barrel, if not completely dominate the in-plane transport, make the largest contribution to it. An alternative explanation would be that the minima exist only in the bulk (but not near the surface as detected by the STM measurements), and live on the outer barrels, as predicted by theory, but disappear near the surface.

To summarize, we have measured the angle-resolved low temperature specific heat in an external in-plane magnetic filed. We find that, as the field direction rotates with respect to the crystallographic axes, fourfold oscillations of the specific heat appear, indicating a strong fourfold anisotropy of the order parameter in $FeSe_{0.45}Te_{0.55}$. The results can be formally interpreted in terms of a nodal $d_{xy}$ gap at the $\Gamma$-FS, but such interpretation is not consistent with other experiments on the same compound. A consistent interpretation can be provided in terms of an order parameter that has deep minima (deeper than roughly 50% of the maximal gap) on the electronic Fermi surfaces, located at the crossing point of the $\Gamma M$ direction (the Fe-Fe bond direction) in the Brillouin zone and the electron pockets. We cannot distinguish between two possibilities, that the minima are located on the outer electron barrels, or on the inner ones. The former one is consistent with existing model calculations, while the latter is consistent with the STM tunneling data on the same compound, and with thermal conductivity.

**Acknowledgements**

We thank D. Agterberg, P. Hirschfeld, K. Kuroki, D. H. Lee, Y. Matsuda, T. Shibauchi and F.

Wang for valuable discussions. This work was supported by the Natural Science Foundation of China, the Ministry of Science and Technology of China (973 Projects No.2006CB601000, No. 2006CB921802), and Chinese Academy of Sciences (Project ITSNEM).

**Competing financial interests**

The authors declare that they have no competing financial interests.

Correspondence and requests for materials should be addressed to Hai-Hu Wen at

hhwen@aphy.iphy.ac.cn

**Figure Legends**

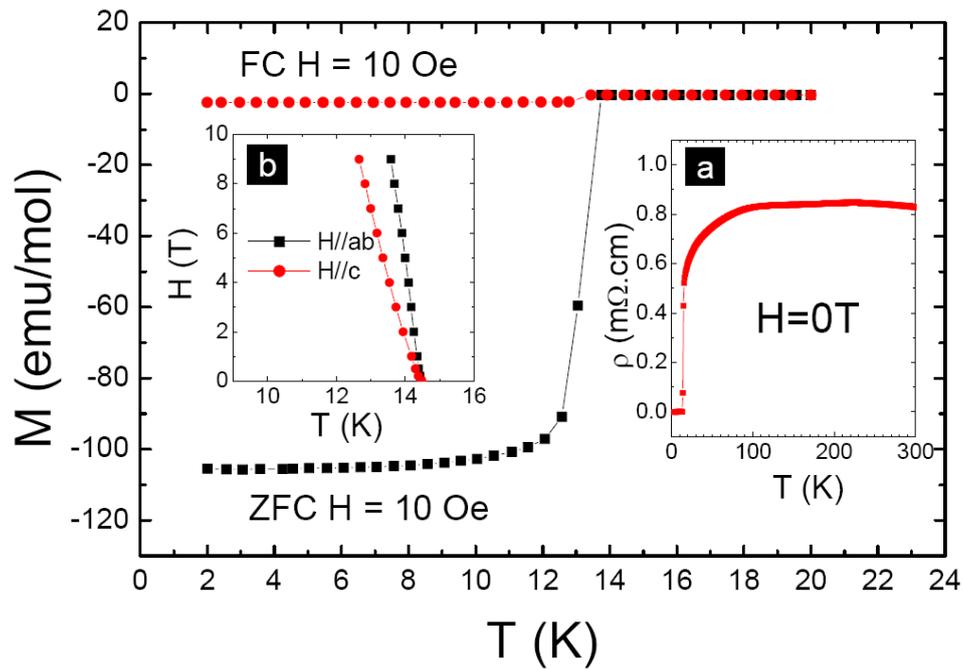

**Figure 1** Main panel: Zero-field-cooling (ZFC) and field-cooling (FC) magnetization measured in the field of 10 Oe. The magnetization in the low temperature region is flat and the Meissner screening volume is almost 100%. The inset (a) shows a sharp resistive transition at $T_{c,mid} \approx 14.5$ K. The inset (b) displays the upper critical fields for in-plane and out-of-plane field orientations.

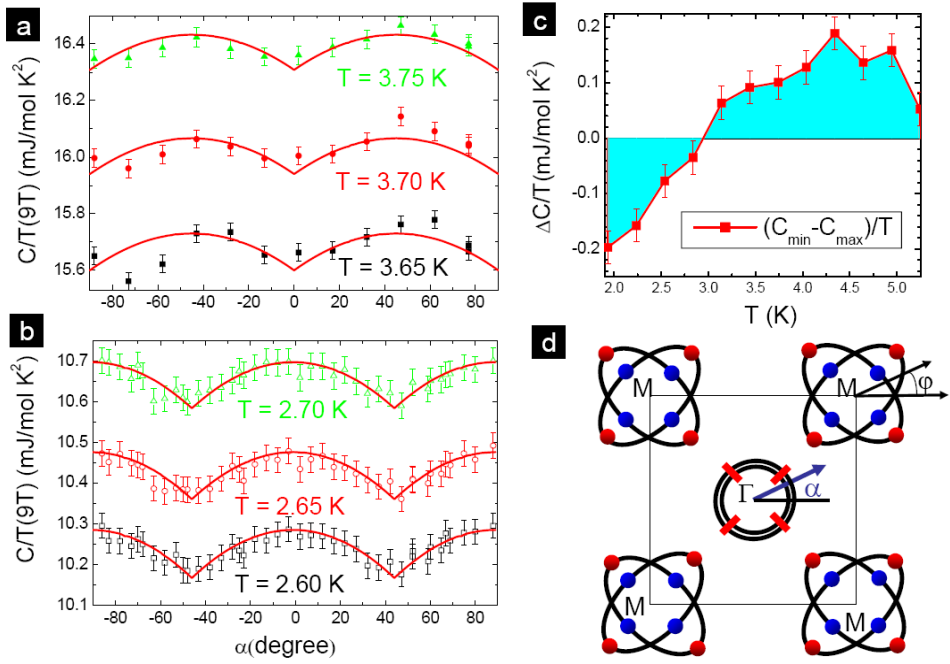

**Figure 2** The angle dependence of the specific heat coefficient at (a) 3.65, 3.7 and 3.75 K and (b) 2.6, 2.65 and 2.7 K at an in-plane magnetic field of 9 T, where the $\alpha$ is the angle enclosed by the external field and the Fe-Se-Fe bond direction. Fourfold oscillations are observed and the amplitude is about 0.12 mJ/mol-$K^2$ (T = 2.6 K). The maximum of C/T locates at about zero degree (H∥Fe-Se-Fe), while the oscillation reverses its sign when the temperature is increased to about 3.7 K. The red lines are theoretical simulations (see Supplementary Material Method-III) using the $d(x^2-y^2)$ order parameter $\Delta_k = \Delta_0(\cos k_x - \cos k_y)$. The actual functional dependence is not important: any reasonable model that yields the gap nodes located at the same directions [42-43] would produce a very similar angular dependence. (c) The temperature dependence of the difference of C/T at 0° and 45°, a crossover was clearly seen at about 2.9 K. (d) The possible nodes or gap minima suggested by our data are located at the points marked by the red and blue balls on the folded electron Fermi pockets, and by red dashes on the hole surfaces. The ellipticity of the electron pockets is exaggerated in this drawing.

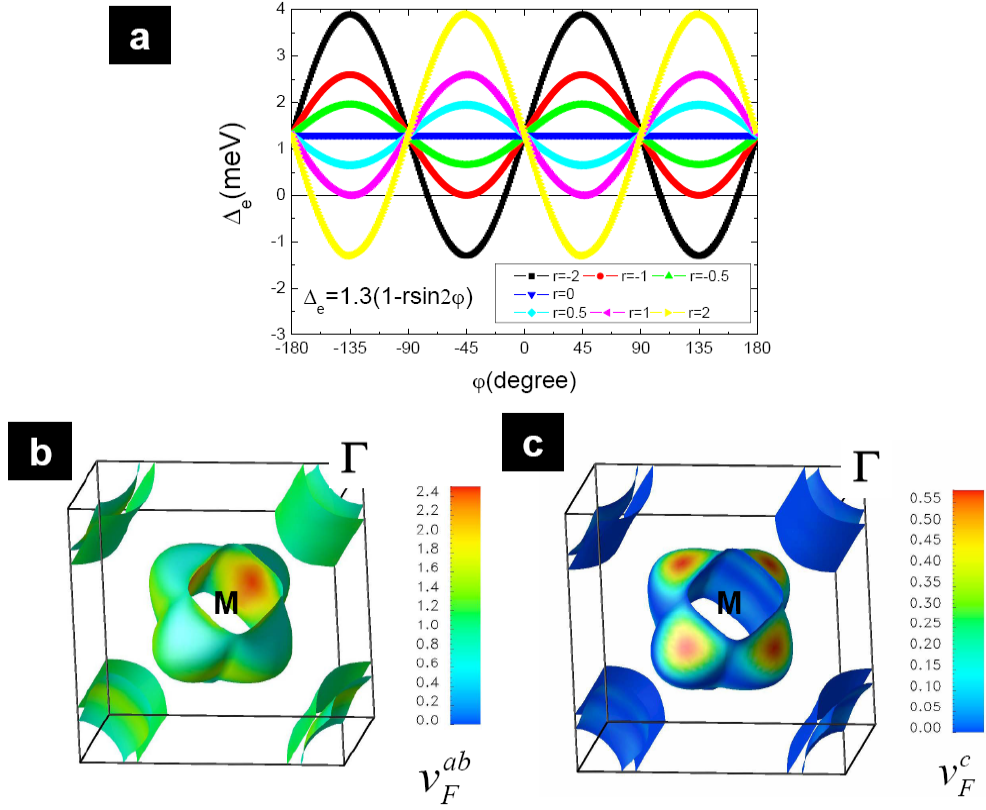

**Figure 3** (a) The angle dependence of the gap on the electron pocket around M($\pi,\pi$) based on the extended s-wave gap function $\Delta_e(\varphi) = \Delta_0[1 - r\sin 2\varphi]$ with $\Delta_0$ = 1.3 meV. The minima are located at 45° (Fe-Fe or ΓM direction) if $1 > r > 0$. For the negative $r$ maxima are located at -45°. For $|r| > 1$ instead of minima, sign-changing nodes appear. (b) and (c): The Fermi surfaces of $FeSe_xTe_{1-x}$, as calculated from first principles. Note strong warping of the outer electron pocket (now shown in the center of the picture for a better view). The coloring represents the function $(v_F^{ab})^2/|v_F|$, which controls the electronic transport in the ab-plane (b), and the same for $(v_F^c)^2/|v_F|$ (c-axis transport) (c).

**Supplementary materials**

**Method-I: Growth and characterizations of the FeSe$_{0.45}$Te$_{0.55}$ single crystals**

The FeSe$_{0.45}$Te$_{0.55}$ single crystals were grown by self-flux method. Powders of Fe, Se and Te were mixed in stoichiometric ratio and filled in a ceramic crucible. The weighing, mixing and pressing procedures were performed in a glove box filled with highly pure Ar gas where both O$_2$ and H$_2$O concentrations were less than 0.1 ppm. The crucible with the starting materials was sealed in an evacuated quartz tube. It was heated up to 720$^{\circ}$C, kept for 10 hrs, and further heated up to 1050$^{\circ}$C for melting the material, then it was slowly cooled down to 720$^{\circ}$C at a rate of 5$^{\circ}$C/hr before the furnace was shut off. The samples used for the present measurements were 2.6 mm × 4.7 mm × 0.35 mm (thickness) in dimension. Well cleaved crystals were sealed in an evacuated quartz tubes again and annealed at 400$^{\circ}$C for more than 300 hrs. All annealed samples show sharp superconducting transitions as shown by the diamagnetic measurements. The resistivity in different magnetic fields are shown in Fig1Sup, one can see that the resistive transition is very sharp and the broadening of resistive transitions are quite narrow, indicating a high upper critical magnetic field.

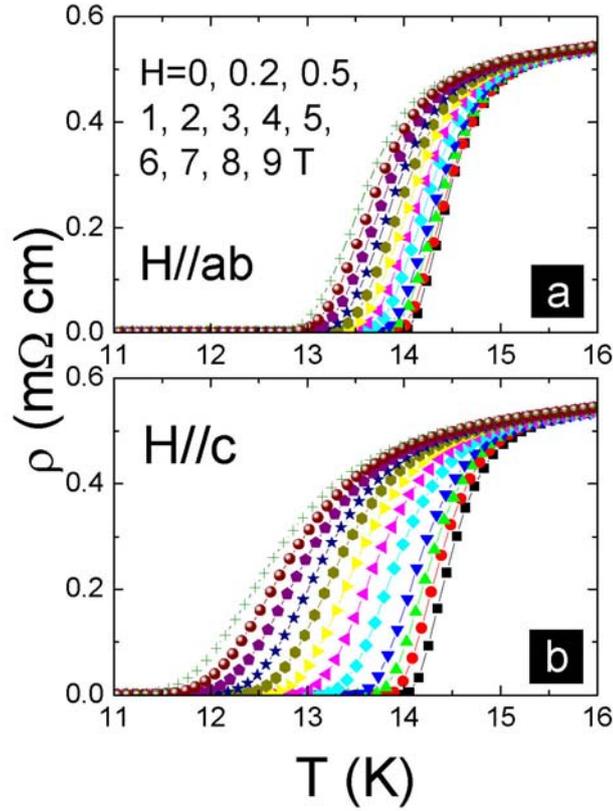

Fig.1sup. Resistivity at magnetic fields up to 9 T is shown in (a) H||ab and (b) H||c. The curves from right to left correspond to the magnetic fields of 0 to 9 T with the values given in (a). It is clear that the resistive broadening is very narrow in the magnetic fields, indicating a high upper critical magnetic field. The anisotropy determined from the slope ratio between $dH_{c2}(T)/dT|_{Tc}$ (H||ab) = -10.4 T/K and $dH_{c2}(T)/dT|_{Tc}$ (H||c) = -5.26 T/K is about 1.98.

**Method-II: Measurements and verifications**

The specific heat was measured using the relaxation technique based on a home made measuring puck which rotates with the sample in the dewar of a Quantum Design instrument PPMS-9T. Due to a careful design, during the rotation, the FeSe planes of the sample are always parallel to the magnetic field. The measurement puck was tested by measuring Nb.

The results were fully consistent with those reported in the literature. During the measurements for each angle the sample was cooled down in a magnetic field, with the angle between the crystallographic axis and the direction of the magnetic field fixed. After the measurement had been finished, the sample was warmed up for changing the magnetic field (all measurements were performed in the field-cooling mode) and rotated to another angle. In order to suppress the noise, the measurements were repeated 300 times for each data point shown in Figure 2(a)-(b). In order to check that the observation of the fourfold oscillations in the magnetic field in $FeSe_{0.45}Te_{0.55}$ was not an artifact, we measured the specific heat at $H = 0$ by rotating the sample $FeSe_{0.4}Te_{0.60}$, and a conventional superconductor, Nb, in an in-plane field $H = 0.4$ T. As shown in Fig2sup, one can see that the data from Nb at $H = 0.4$ T do not present any discernable oscillations. The same happens for the sample $FeSe_{0.4}Te_{0.60}$ at $H = 0$ and $T = 2.6$ K (data not shown here).

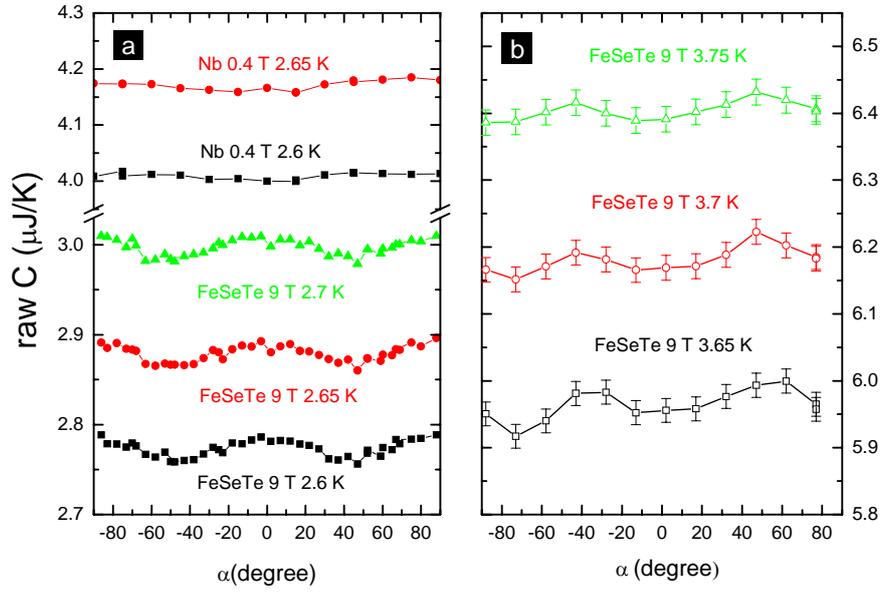

Fig.2sup. Raw data of the angle dependence of the measured specific heat for $FeSe_{0.45}Fe_{0.55}$ in the in-plane fields of 9 T, and Nb at H = 0.4 T. It is clear that the fourfold oscillations are only observed in H = 9 T for $FeSe_{0.45}Te_{0.55}$, while absent in the case of Nb (H = 0.4 T).

In order to analyze the different contributions of the specific heat of the $FeSe_{0.45}Te_{0.55}$, in Fig.3sup we present the specific heat coefficient $\gamma = C/T$ measured at H = 0 and 9 T (H||ab) in wide temperature region. It is clear that the specific heat anomaly at about 12 K (middle point of the SH anomaly) is rather sharp. The low temperature data at H = 0 T has a slight upturn, which is given by the Schottky anomaly. Although a linear extrapolation of the data in the high temperature region can give a reasonable assessment of all terms, we shall still do the global fitting for the data measured at H = 0. The total specific heat $C_{total}$ includes the residual term $\gamma_0 T$, the phonon part $C_{ph}/T = \beta T^2$, the Schottky anomaly part $\gamma_{Sch}$. The specific heat resulting from the superconducting part is assumed to be negligible in this low temperature region. Therefore we have

$$C_{total}/T = \gamma_0 + C_{ph}/T + \gamma_{Sch} = \gamma_0 + \beta T^2 + n\left(\frac{g\mu_B H_{eff}}{k_B T}\right)^2 \frac{\exp(g\mu_B H_{eff}/k_B T)}{[1+\exp(g\mu_B H_{eff}/k_B T)]^2}, \quad (1\text{sup})$$

where g is the g-factor, taking 2 for the case s = 1/2 and assuming weak spin-orbital coupling, $H_{eff}$ is the effective crystal field, n is a prefactor related to the number of paramagnetic centers. Using the above equation, we get a good fit to the data at H = 0, yielding $\gamma_0$ = 1.66 mJ/mol-K$^2$, β = 0.924 mJ/mol-K$^4$, n = 21 mJ/mol-K$^2$, $H_{eff}$ = 3.1 T. As shown in the inset of Fig3sup, after removing the Schottky anomaly, the data exhibit roughly linear behavior, extrapolating to $\gamma_0$ at T = 0 K. No upturn in H = 9 T is observed, indicating that the Schottky anomaly is weak. Based on above equation, the Schottky term in our experiment at H = 9 T is much smaller than that in H = 0 T in the temperature region of 2.6-2.7 K, in which we collected the ARSH data. Even if the Schottky anomaly is present, it should not give rise to any particular angular dependence.

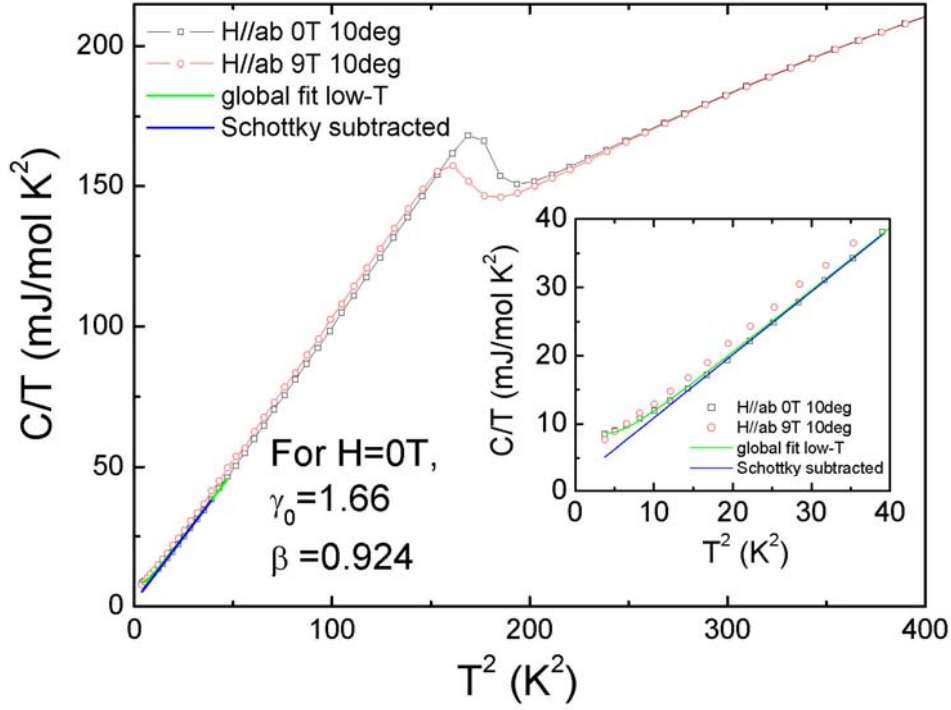

Fig.3sup. The specific heat in H = 0 and 9 T (H||ab). A slight upturn can be seen at low fields, induced by the Schottky anomaly. The data can be well fitted globally by a single formula (eq.1sup) (green solid lines). The data after removing the Schottky anomaly is shown by the blue line.

**Method-III. Simulations for $d(x^2-y^2)$-wave gap structure**

In this Section we consider two representative examples: $d(x^2-y^2)$-wave type nodes on the hole FS ($d_{xy}$) and accidental nodes within an extended $s$ model on the electron FS. In both cases we used the simplified unfolded Brillouin zone and unfolded Fermi surfaces (Fig. 4sup). For the former case, we used the gap function $\Delta_k = \Delta_0(\cos k_x - \cos k_y)$. For the latter case we use $\Delta_k = \Delta_0 \cos k_x \cos k_y$ (corresponding to the gap on the electron pockets: $\Delta_e = \Delta_0(1 - r\sin 2\varphi)$ with $r \approx 1$). The latter function, as shown at Fig. 4sup, has nodes or the gap minimum along the ΓM or Fe-Fe bond directions on the electron pockets.

In the $d(x^2-y^2)$-wave case we can rewrite Eq. 2 as

$$N_0 = N_M + N_\Gamma^0 \int_0^{2\pi} \frac{d\varphi}{2\pi} \min\left[1, \frac{A}{\cos^2 2\varphi} \sin^2(\varphi - \alpha)\right]$$

(2sup)

where $A = 2a^2\hbar^2 v_F^2 H / \pi\Phi_0\Delta_0^2$ [29]. The integral is taken over the hole FS only (assuming full isotropic gaps on the electron pockets), since only this FS has nodes in this case. Taking $v_F = 1\times10^5$ m/s as discussed in the main text, $\Delta_0 = 4$ meV [7], and H = 9 T, we get A = 0.27. The resulting simulation is shown by red lines in Figures 2(a) and (b). Obviously, this model can describe the data very well. Less trivial case is the extended s-wave because the gap minima or nodes are dependent on the anisotropic parameter r. Since our data indicates the gap minima, not nodes, along the ΓM or Fe-Fe bonds direction, we leave the calculation on the quasiparticle density of states based on this model to the future work.

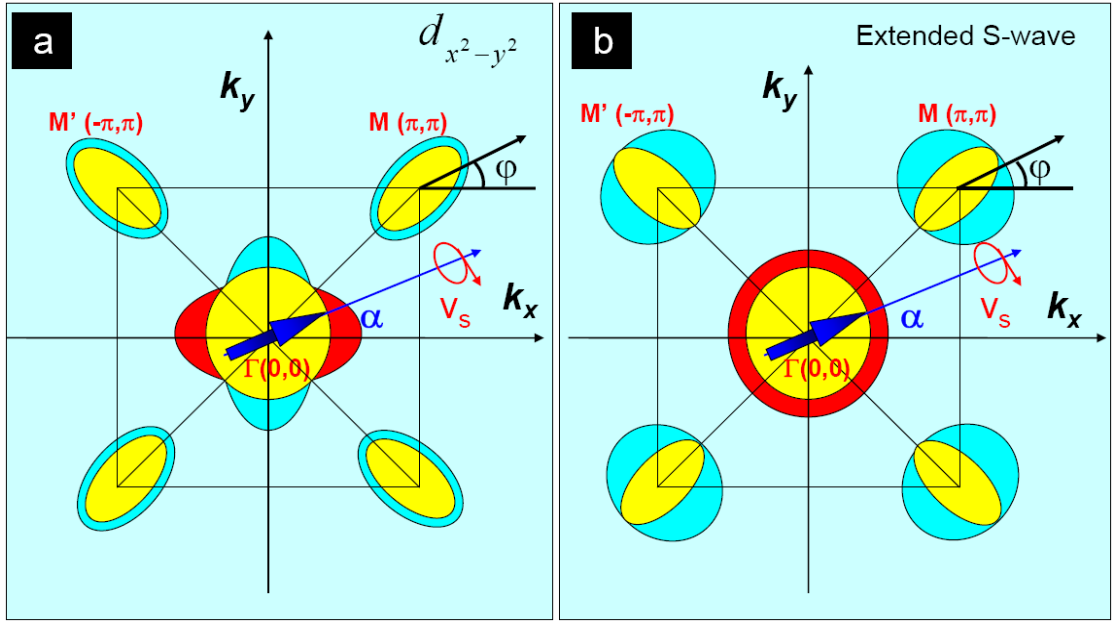

**Figure 4sup.** Cartoon representations of the Fermi surfaces and the gap functions in the unfolded zone (one Fe unit cell) for (a) the d-wave gap: $\Delta_k = \Delta_0(\cos k_x - \cos k_y)$ and (b) the extended s-wave gap $\Delta_k = \Delta_0 \cos k_x \cos k_y$ (corresponding to the gap on the electron pockets:

$\Delta_e = \Delta_0(1 - r\sin 2\varphi)$ with r ≈ 1). The yellow shapes represent the Fermi surfaces at Γ(0,0) and M(±π,±π). The red and the cyan areas represent the positive and negative order parameters. The blue arrow indicates the magnetic field direction with the red circle and arrow of the flowing direction of the supercurrent of the vortices. The angle between the field and the (1,0) direction is denoted by α. The winding angle on the electron pocket is denoted by φ. The ellipticity of the electron pockets is exaggerated.

Finally, we can, using our simulations described above, get a quantitative assessment of our data, as a consistency check. Experimentally, we found that the fourfold oscillations have an amplitude of about $\Delta C/T = 0.12 mJ/mol-K^2$. The total specific heat coefficient at H = 9 T can be written as

$$C/T = \gamma_0 + C_{ph}/T + (\gamma_e^h + \gamma_e^e), \qquad (4sup)$$

where $C_{ph}/T$ is the contribution from phonons, $\gamma_0$ is the residual term of specific heat at H = 0 T at T = 0 K, $\gamma_e^h$ and $\gamma_e^e$ are the electronic contributions from the Γ-FS and the M-FS at 9 T respectively. In Methods-II, we presented the specific heat of the sample at H=0 and 9 T with H∥ab plane. One can see that the specific jump near $T_C$ is very sharp, indicating a good quality of the sample. Nevertheless, even for this sample we still see a sizable residual term $\gamma_0$ at H=0 and T=0. This residual term $\gamma_0$ can be partially attributed to the non-superconducting fraction, and partially to the impurity scattering at the gap nodes (or gap minimum). There is a slight upturn in the data C/T vs. $T^2$ at H = 0 T, which is attributed to the Schottky anomaly. Using a global fit to the data at H = 0, as described in the Methods-II section, we get $\gamma_0$ = 1.66 mJ/mol-$K^2$ and β = 0.924 mJ/mol-$K^4$, therefore we have $(\gamma_e^h + \gamma_e^e) \approx 3 mJ/mol-K^2$ at 9 T. We can fit our data to Eq.4sup and extract the parameters $\gamma_e^h$ and $\gamma_e^e$ with the d-wave pairing model (using Eq.2sup by taking A = 0.27). The derived parameters are listed as below

Table-I: The electronic contributions of specific heat $\gamma_e^h$ and $\gamma_e^e$ at 9 T (in unit of mJ/mol-K$^2$) coming from different pockets, as extracted from fitting the data to the d-wave gap function.

| Gap function → <br> Temperature ↓ | $\Delta_k = \Delta_0(\cos k_x - \cos k_y)$ |
|---|---|
| T = 2.60 K | $\gamma_e^h$=0.97, $\gamma_e^e$=1.944 |
| T = 2.65 K | $\gamma_e^h$=0.95, $\gamma_e^e$=1.901 |
| T = 2.7 K | $\gamma_e^h$=0.93, $\gamma_e^e$=1.884 |